\documentclass[a4paper,11pt]{article}
\usepackage{pos}
\usepackage{amsmath}   
\usepackage{hyperref}

\makeatletter
\renewcommand{\printHeadAuthors}{}
\makeatother

\title{Towards Unveiling the Origins of the Milky Way
Bulge through Multi-band-Messenger Sky Surveys}

\author*[a,e]{Hai-Feng Wang}
\author[b]{Xiao Han}
\author[a]{Giovanni Carraro}
\author[c]{Mart\'in L\'opez-Corredoira}
\author[d]{Yuan-Sen Ting}
\author[a]{Guan-Yu Wang}

\affiliation[a]{Dipartimento di Fisica e Astronomia “Galileo Galilei”, Universit\`a degli Studi di Padova, Vicolo Osservatorio 3, I-35122, Padova, Italy}
\affiliation[b]{Wuhan University, School of Physics and Technology, Bayi Road 299, Wuhan, P.\,R.\,China}
\affiliation[c]{Instituto de Astrof\'\i sica de Canarias, E-38205 La Laguna, Tenerife, Spain}
\affiliation[d]{Department of Astronomy, The Ohio State University, Columbus, OH 43210, USA}
\affiliation[e]{Local Universe and Time-Domain Astronomy Laboratory (LUTDLab), China West Normal University, Nanchong, 637002, China}

\emailAdd{haifeng.wang.astro@gmail.com}

\abstract{We analyze the structure and chemo-dynamical properties of the Galactic bulge using ab-type RR Lyrae stars (RRabs) from OGLE-IV and giant stars from APOGEE and Gaia. Orbital integration of 1,879 RRab variables reveals three populations: central bulge, inner bulge, and halo/disk contaminants. Inner bulge RRabs display bar-like kinematics, whereas central bulge stars show slower rotation and lower dispersion. APOGEE data for 28,188 stars confirm these kinematic trends and reveal a bimodal chemical distribution, indicating distinct formation pathways. Our results support a pseudo-bulge origin of the inner bulge through disk instability, with the overall morphology better described as boxy rather than X-shaped. Through the integration of multi-messenger, multi-band data, our collaboration aims to provide deeper insights into the physical properties and evolutionary history of the Galactic bulge.}

\FullConference{Multifrequency Behaviour of High Energy Cosmic Sources - XV (MULTIF2025)\\
 9-14 June, 2025\\
Mondello, Palermo, Italy\\}

\begin{document}
\maketitle

\section{Introduction}

The Galactic bulge hosts multiple stellar populations with distinct structural, kinematic, and chemical signatures. These are commonly interpreted as arising from early gas collapse and mergers that produced a classical bulge, together with secular evolution driven by disk instabilities that produced a pseudo-bulge; the coexistence of both components remains uncertain \cite{2020AJ....159..270K,2018ARA&A..56..223B,2016PASA...33...26B,2013ApJ...763...26O,2008A&A...486..177Z}.
Metallicity has proven useful for distinguishing populations, as metal-rich stars tend to rotate faster with lower dispersion, while metal-poor stars show hotter kinematics and central concentration \cite{2017A&A...599A..12Z, 2020MNRAS.498.5629D, 2021MNRAS.501.5981L, 2021A&A...656A.156Q}. However, not all stars toward the bulge are genuine members; some are halo or disk interlopers. Recent studies show that orbital analysis—particularly apocenter-based classification of RRLs—often outperforms metallicity as a discriminator \cite{2020AJ....159..270K, 2024A&A...687A.312O}.
Despite extensive datasets, many aspects of the Galactic bulge remain uncertain, including the roles of different progenitors, secular evolution, the origin of ancient metal-poor populations \cite{2022ApJ...941...45R}, the X-shaped structure \cite{2012ApJ...757L...7L}, and the identification of distinct populations such as Kraken and Aurora \cite{2024MNRAS.528.3198B}. Using OGLE-IV and APOGEE DR17 data, we investigate the chemo-dynamical properties of stars in the central and inner regions of the Galactic bulge, aiming to explore their stellar classification and to study their formation and evolutionary history.

\section{Data}
\subsection{RRabs or RRLs of OGLE survey}
\label{section:data}

The dataset comprises 27,258 fundamental-mode RRab stars from the OGLE-IV catalog \cite{2014AcA....64..177S}, providing photometry in the $I$- and $V$-bands, positions, periods, and other parameters. Following Pietrukowicz et al.\cite{2015ApJ...811..113P}, we cleaned the sample and derived metallicities ([Fe/H]) and distances using the RRab period-luminosity relation, resulting in 17,817 high-quality stars.
Through cross-matching the RRL sample with Kunder et al.\cite{2020AJ....159..270K} for line-of-sight (LOS) velocities and Gaia EDR3 for proper motions, we obtained a final sample of 1,879 stars with both kinematic measurements. Figure~\ref{lbscatter_disthist} shows the spatial distribution of the 1,879 RRabs in $(l, b)$ plane (yellow points, left) and their distances from the Sun (orange histogram, right).

We adopt a Galactocentric Cartesian system with X toward the Sun, Y along Galactic longitude $l = 90^\circ$, and Z toward the North Galactic Pole, and use GALPY \cite{2015ApJS..216...29B} to derive the three-dimensional spatial distribution of the RRab stars.

\subsection{RGB and RC Stars of APOGEE survey} 
\label{subsetction:RGB and RC Stars}

We use astrometric and spectroscopic data from \textsc{APOGEE} DR17 \cite{2022ApJS..259...35A}, including LOS velocities and chemical abundances. Distances and extinctions are adopted from previous work \cite{2023A&A...673A.155Q}, derived with \textsc{StarHorse} \cite{2016A&A...585A..42S,2018MNRAS.476.2556Q,2019A&A...628A..94A} for over 10 million stars. These distances are model-dependent and may vary with a different stellar density distribution along the line of sight.
By cross-matching \textsc{APOGEE} LOS velocities, Gaia DR3 proper motions, and \textsc{StarHorse} distances, we computed Galactocentric Cartesian and cylindrical velocities using GALPY \cite{2015ApJS..216...29B}, with mean uncertainties of 0.13 km s$^{-1}$ (LOS) and 0.16 mas yr$^{-1}$ (PMs). We selected stars within $|\mathrm{X}_{\mathrm{Gal}}| < 5$ kpc, $|\mathrm{Y}_{\mathrm{Gal}}| < 3.5$ kpc, $|\mathrm{Z}_{\mathrm{Gal}}| < 1$ kpc, $|l| < 15^\circ$, and $|b| < 15^\circ$, yielding a final sample of 28,188 predominantly RGB and RC stars \cite{2021A&A...656A.156Q}.

\begin{figure*}[!t]
  \centering
  \includegraphics[width=5in]{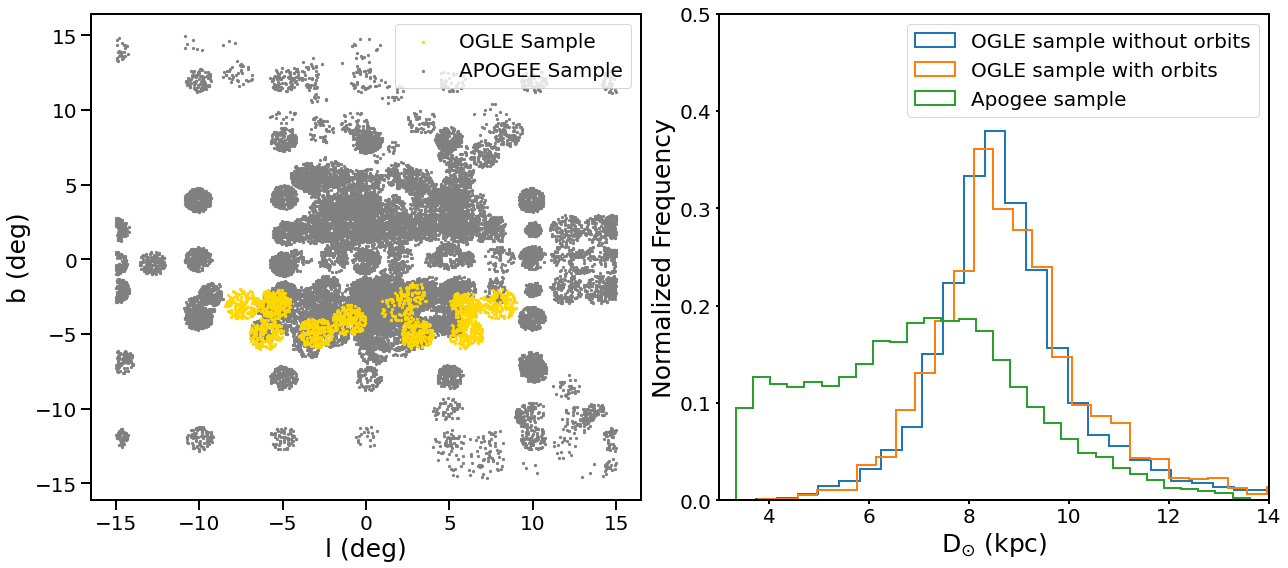}
  \caption{Left: Galactic positions of 1,879 RRab stars with orbits (yellow) and 28,188 APOGEE stars (gray). Right: Heliocentric distance histograms for the OGLE and APOGEE samples: 17,817 RRab stars before orbital selection (blue), 1,879 RRab stars after orbital selection (orange), and 28,188 APOGEE stars (green).}
  \label{lbscatter_disthist}
\end{figure*}

The spatial distribution of the sample in the $(l, b)$ plane is shown as gray points in the left panel of Fig.\ref{lbscatter_disthist}, while the right panel displays the heliocentric distance histogram (green) of the \textsc{APOGEE} subsample.

\begin{figure*}[!ht]
  \centering
  \includegraphics[width=4.5in]{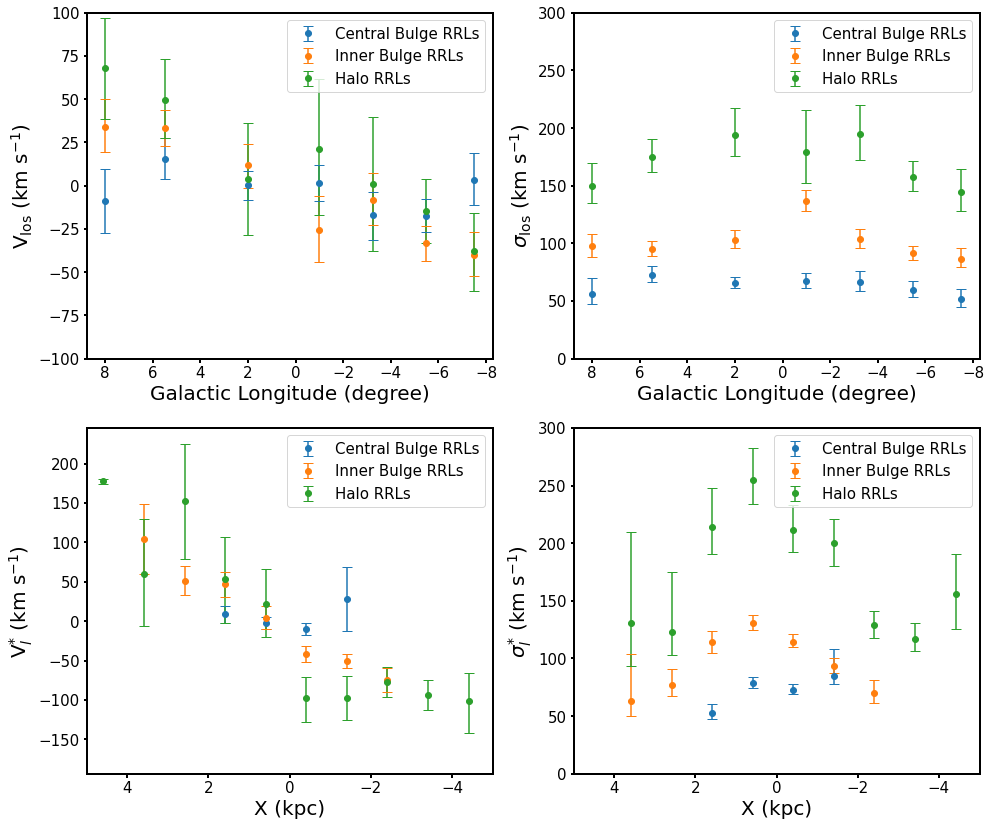}
  \caption{Top: The mean LOS velocity and velocity dispersion maps of different stellar populations as a function of Galactic longitude. Blue points represent central bulge RRabs (\( r_{\text{apo}} < 1.8 \, \text{kpc} \)), orange points represent inner bulge RRabs (\( 1.8 \, \text{kpc} \leq r_{\text{apo}} < 3.5 \, \text{kpc} \)), and green points represent halo interlopers (\( r_{\text{apo}} \geq 3.5 \, \text{kpc} \)). Bottom: The mean $v^{*}_{l}$ and velocity dispersion maps for the same stars.}
  \label{Orbit_rv_tv}
\end{figure*}

\begin{figure*}[!ht]
  \centering
  \includegraphics[width=4.5in]{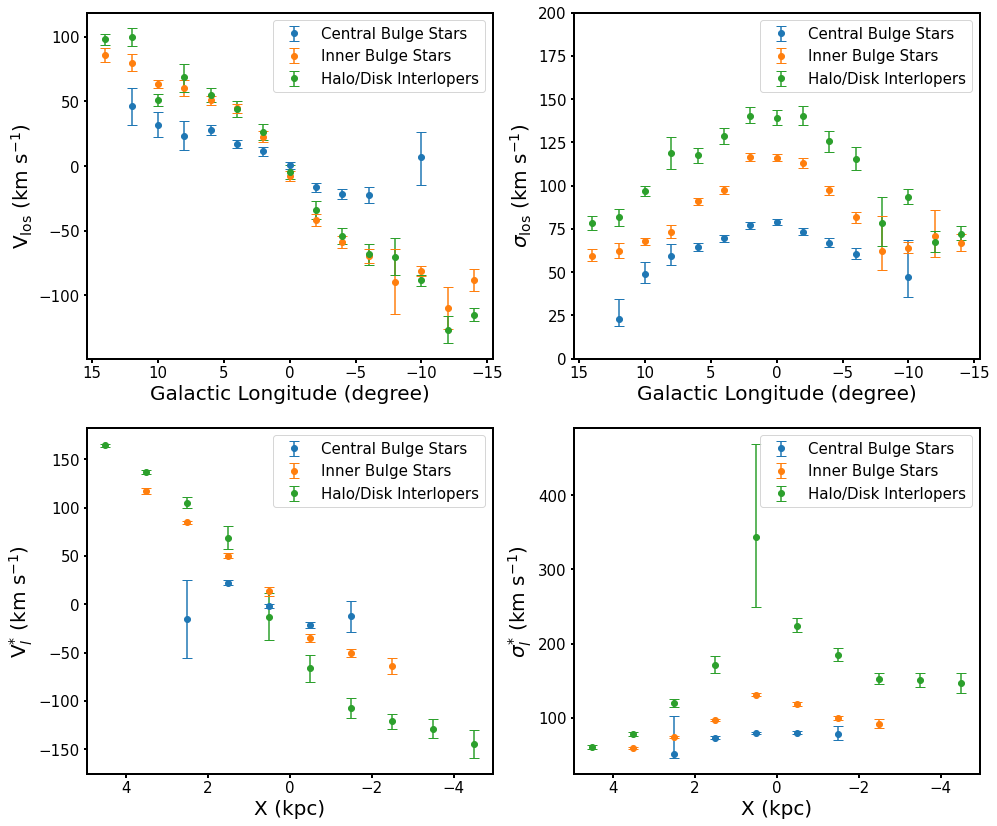}
  \caption{Top: The mean LOS velocity and velocity dispersion maps of different stellar populations as a function of Galactic longitude. Blue points represent central bulge RGBs and RCs (\( r_{\text{apo}} < 1.8 \, \text{kpc} \)), orange points represent inner bulge RGBs and RCs (\( 1.8 \, \text{kpc} \leq r_{\text{apo}} < 3.5 \, \text{kpc} \)), and green points represent halo/disk interlopers (\( r_{\text{apo}} \geq 3.5 \, \text{kpc} \)). Bottom: The mean $v^{*}_{l}$ and velocity dispersion maps for the same stars.}
  \label{lonVlos_XVll_subsetABC}
\end{figure*}

\begin{figure*}[!ht]
  \centering
  \includegraphics[width=5in]{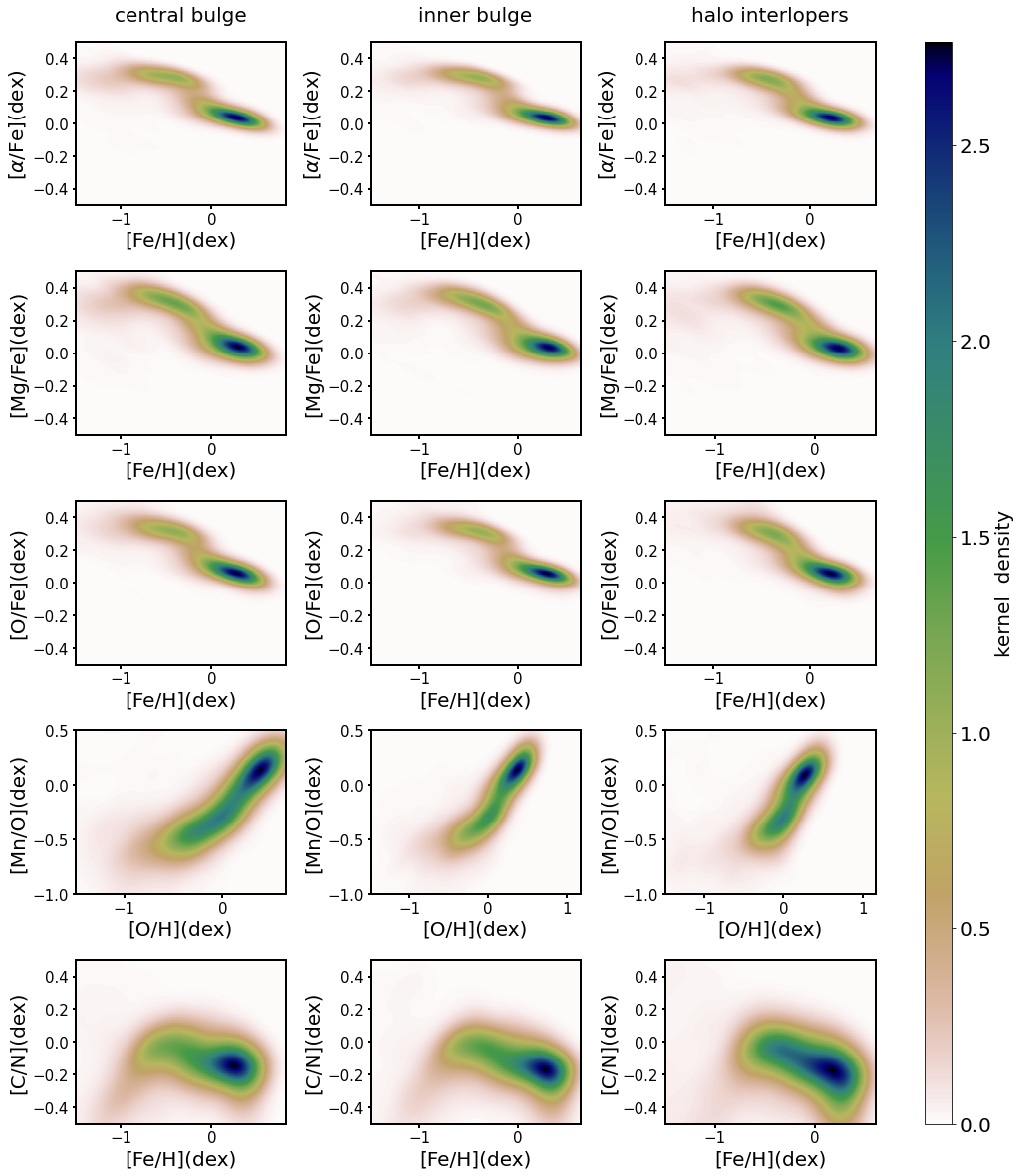}
  \caption{Two-dimensional chemical abundance plots of APOGEE RGBs and RCs in the central bulge (left), inner bulge (middle), and halo/disk interlopers (right). From top to bottom, respectively, are [$\alpha$/Fe] vs. [Fe/H], [Mg/Fe] vs. [Fe/H], [O/Fe] vs. [Fe/H], [Mn/O] vs. [O/H], and [C/N] vs. [Fe/H]. The [Mn/O] patterns suggest different star formation histories for the inner bulge and central bar.}
  \label{2DMetallicity_subsetABC}
\end{figure*}

\begin{figure*}[!ht]
  \centering
  \includegraphics[width=4.5in]{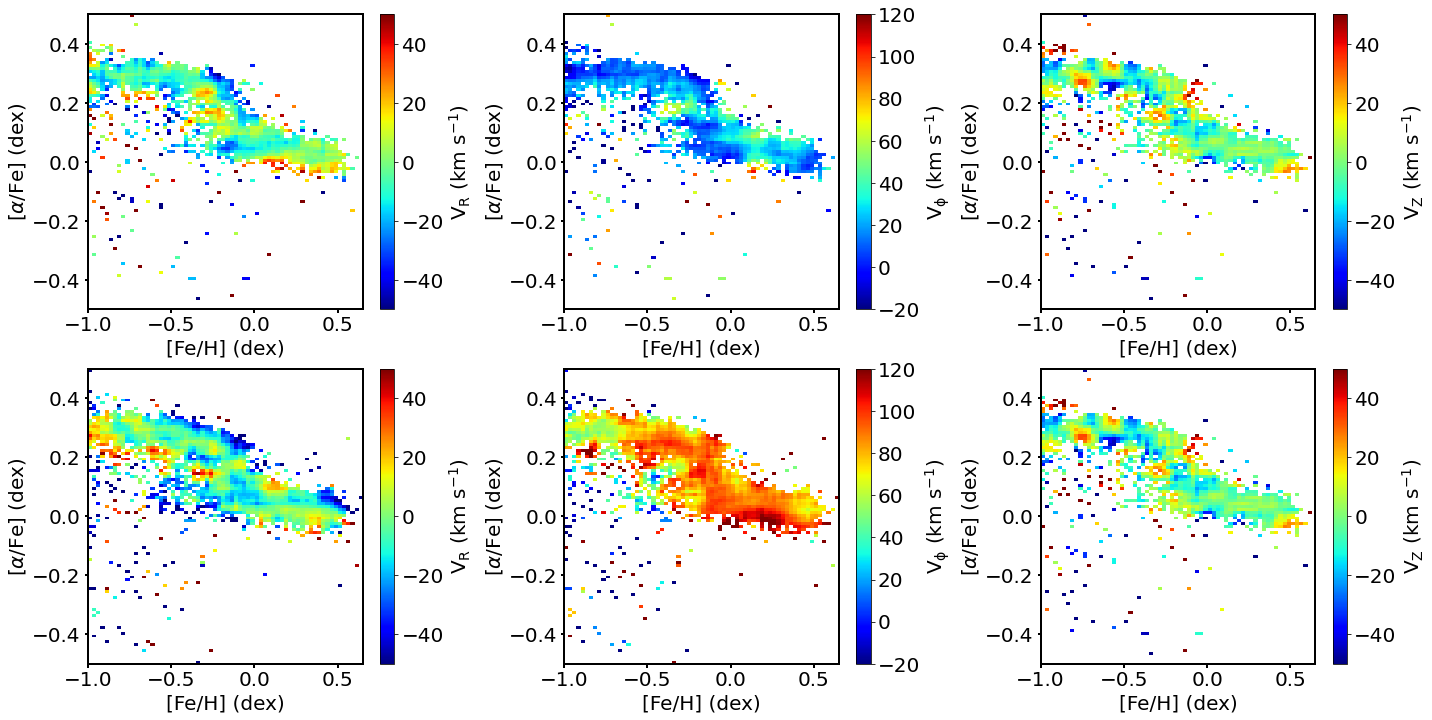}
  \caption{Distributions of stars in the [$\mathrm{Fe/H}$]–[$\alpha/\mathrm{Fe}$] plane, color-coded by their cylindrical velocity components: radial velocity ($\mathrm{V_R}$, left column), azimuthal velocity ($\mathrm{V_\phi}$, middle column), and vertical velocity ($\mathrm{V_Z}$, right column). The top row corresponds to stars located in the central bulge region, while the bottom row shows stars in the inner bulge.}
  \label{FeH_aFe_Vcyl}
\end{figure*}

\section{Orbits}
\label{section:Orbits}

Due to the overlap of bulge, disk, and halo stars near the Galactic center, selecting a pure bulge sample based solely on distance and position is infeasible. We therefore performed orbital analysis to distinguish stars confined to the bulge from halo and disk interlopers. 

Orbits were integrated using \textsc{GALPY} \cite{2015ApJS..216...29B} with the \texttt{MWPotential2014} model, comprising a PowerSphericalPotentialwCutoff bulge, a Miyamoto-Nagai disk \cite{1975PASJ...27..533M}, and a Navarro-Frenk-White halo \cite{1997ApJ...490..493N}, together with a KeplerPotential for the central supermassive black hole and a DehnenBarPotential for the Galactic bar \cite{2000AJ....119..800D}. We assumed a circular velocity of $220~\mathrm{km\,s^{-1}}$ \cite{2012ApJ...759..131B} and a Galactic center distance of 8.277~kpc \cite{2022A&A...657L..12G}, and integrated orbits backward for 5~Gyr.
Using the apogalactic distance ($r_{\rm apo}$) derived from the orbital parameters, we classified stars following Kunder\cite{2022Univ....8..206K}: those with $r_{\rm apo} < 3.5~\mathrm{kpc}$ were considered bulge stars, subdivided into central bulge stars ($r_{\rm apo} < 1.8~\mathrm{kpc}$) and inner bulge ($1.8 \le r_{\rm apo} < 3.5~\mathrm{kpc}$) stars, while stars with $r_{\rm apo} \ge 3.5~\mathrm{kpc}$ were treated as halo interlopers.

\section{Results}
\label{section:Results}
\subsection{Kinematics of RRabs from OGLE}
\label{subsection:kinematics of RRabs from OGLE}

We analyzed the kinematics of each stellar population. Figure~\ref{Orbit_rv_tv} shows the LOS velocity (top) and transverse velocities ($v^{*}_{l}$) (bottom) distributions for central bulge RRLs, inner bulge RRLs, and halo interlopers, along with their intrinsic velocity dispersions. The inner bulge RRLs exhibit clear rotation in both LOS and $v^{*}_{l}$, whereas central bulge RRLss show negligible net rotation. The intrinsic dispersions indicate that the central bulge has the lowest velocity dispersion, followed by the inner bulge, with halo interlopers showing the highest dispersion.

\subsection{Kinematics and Chemistry of APOGEE stars}
\label{subsection:Kinematics and Chemistry of RGBs and RCs}

The kinematics of APOGEE stars were analyzed in the same manner as the RRLs. Their LOS velocities and $v_{l}^{*}$, shown in Figure~\ref{lonVlos_XVll_subsetABC}, reveal that central bulge stars rotate more slowly and have smaller velocity dispersions than inner bulge stars, with high-dispersion stars primarily being disk/halo interlopers. These results are consistent with the RRL kinematics, indicating that inner bulge stars trace a bar-like structure with disk-like orbits, whereas central bulge stars do not \cite{2020AJ....159..270K}. In addition, both the RRab and APOGEE stellar samples demonstrate that the inner bulge contains the largest fraction of stars. Taken together, the dominance of the inner-bulge population and its disk-like kinematics strongly support a scenario in which secular evolution is the primary formation channel of the Galactic bulge.

Figure~\ref{2DMetallicity_subsetABC} shows kernel density distributions of stars in the central bulge, inner bulge, and disk/halo interlopers across [$\alpha$/Fe] vs. [Fe/H], [Mg/Fe] vs. [Fe/H], [O/Fe] vs. [Fe/H], [Mn/O] vs. [O/H], and [C/N] vs. [Fe/H] planes. All populations exhibit bimodal distributions, distinguishing an $\alpha$-rich, metal-poor component formed during early Type~II supernova--dominated star formation, and an $\alpha$-poor, metal-rich component influenced by later Type~Ia contributions. Trends in [Mn/O] and [C/N] further reflect supernova enrichment history and stellar evolutionary effects, with bulge populations showing systematic patterns and disk/halo interlopers a more scattered distribution. These results indicate complex, multi-phase star formation and chemical evolution in the Galactic bulge.

Figure~\ref{FeH_aFe_Vcyl} presents the distribution of stars in the [$\mathrm{Fe/H}$]–[$\alpha/\mathrm{Fe}$] plane, with color coding representing the cylindrical velocity components: radial velocity ($\mathrm{V_R}$, left panels), azimuthal velocity ($\mathrm{V_\phi}$, middle panels), and vertical velocity ($\mathrm{V_Z}$, right panels). The upper row corresponds to stars in the central bulge, while the lower row shows stars in the inner bulge. In both regions, the stellar populations exhibit a clear bimodality, with one component being metal-poor and $\alpha$-enhanced, and the other metal-rich with lower $\alpha$ abundances. Notably, the inner bulge stars display significantly higher $\mathrm{V_\phi}$ compared to those in the central bulge, indicating a dynamically distinct component.

\subsection{Boxy and X Shape Bulge Probability}
\label{Fits of the density}

To probe the Galactic bulge structure, we compared observed stellar densities with theoretical models, fitting both the boxy \cite{2005A&A...439..107L} and X-shaped \cite{2013MNRAS.435.1874W} bulge models using the parametrization of Leung et al.\cite{2017ApJ...836..218L}. The boxy model exhibits a symmetric, square-like density distribution, while the X-shaped model shows a double-peaked structure along the line of sight. Least-squares fitting was performed, minimizing the reduced chi-square
\begin{equation}
\chi^2_r = \frac{1}{N-1} \sum_{i=1}^{N} \frac{\left( \rho_{i,\text{model}} - \rho_{i,\text{data}} \right)^2}{\sigma_i^2},
\end{equation}
with only the amplitude treated as a free parameter. Both RRab and APOGEE samples consistently favor the boxy model ($\chi^2_r \approx 0.99$, $p \sim 53\%-58\%$) over the X-shaped model ($\chi^2_r \approx 1.00$, $p \sim 42\%-47\%$), with similar results when restricting to inner and central bulge stars. These results indicate that the Galactic bulge is better described by a boxy density distribution.

\section{Summary and conclusions}

In this study, we combined RRab stars from the OGLE-IV survey with Gaia EDR3 proper motions and LOS velocities from Kunder et al.\cite{2020AJ....159..270K} to derive three-dimensional kinematics for 1,879 RRabs. Using apocentric distances, halo interlopers ($r_\text{apo} \geq 3.5$ kpc) were excluded, and the remaining stars were classified as inner bulge ($1.8 \leq r_\text{apo} < 3.5$ kpc) and central bulge ($r_\text{apo} < 1.8$ kpc). Inner bulge RRabs show kinematics consistent with the Galactic bar, while central bulge RRabs exhibit slower rotation and lower velocity dispersion, indicating they do not trace the bar.

Similarly, we analyzed 28,188 stars from APOGEE DR17, combined with Gaia DR3 proper motions and StarHorse distances, primarily consisting of RGB and RC stars. Using the same orbital criteria, stars were classified into central bulge, inner bulge, and disk/halo interlopers. Inner bulge stars align kinematically with the bar, central bulge stars rotate more slowly and show lower velocity dispersion, and disk/halo interlopers have higher dispersion. Chemical abundance analysis reveals bimodal distributions in [$\alpha$/Fe], [C/N], and [Mn/O].

Overall, our results from both RRabs and APOGEE stars confirm a structural division of the bulge into an inner, bar-like component and a central component not associated with the bar. The inner bulge, containing the majority of stars on disk-like orbits, traces the bar formed via secular disk evolution \cite{2010ApJ...720L..72S}, suggesting that secular processes dominate bulge formation. Disk instability or oscillations beyond the inner disk could also be present in some previous works \cite{2019ApJ...877L...7W,2020ApJ...902...70W}. The boxy bulge population appears more prominent than the X-shaped component. Future studies will require more precise distances, stellar ages, and detailed information on accretion events, such as Kraken and GSE, to further disentangle the contributions of secular evolution and spheroid components.

\bibliographystyle{JHEP}       
\bibliography{mybib}           

\end{document}